\documentclass[12pt]{article}
\usepackage{amssymb,graphicx}
\usepackage{epstopdf}
\usepackage{amsmath,amsfonts}
\usepackage{epsfig}

\def\d{\partial}

\newcommand{\be}{\begin{equation}}
\newcommand{\ee}{\end{equation}}
\newcommand{\bea}{\begin{eqnarray}}
\newcommand{\eea}{\end{eqnarray}}
\newcommand{\bg}{\begin{gather}}
\newcommand{\eg}{\end{gather}}
\newcommand{\bseq}{\begin{subequations}}
\newcommand{\eseq}{\end{subequations}}

\newcommand{\pd}[0]{\partial}

\DeclareMathOperator{\diag}{diag}
\begin{document}
\begin{flushright}
\end{flushright}
\begin{center}
  {\LARGE \bf Superluminality \\ in the Fierz--Pauli massive gravity } \\
\vspace{20pt}
M.~Osipov$^{a,b}$, V.~Rubakov$^{a}$\\
\vspace{15pt}
  $^a$\textit{
Moscow Institute of Physics and Technology,\\
Institutskii per., 9, Dolgoprudny, 141700 Moscow Region, Russia
  }\\
\vspace{5pt}
$^b$\textit{
Institute for Nuclear Research of
         the Russian Academy of Sciences,\\  60th October Anniversary
  Prospect, 7a, 117312 Moscow, Russia}\\
    \end{center}
    \vspace{5pt}

\begin{abstract}
We study the propagation of helicity-1 gravitons in the Fierz--Pauli massive
gravity in nearly Minkowski backgrounds. We show that,
generically,
there exist backgrounds consistent with field equations, in which the
propagation is superluminal. The relevant distances are much longer than
the ultraviolet cutoff length inherent in the Fierz--Pauli gravity, so 
superluminality occurs within the domain of validity of the effective
low energy theory. 
There remains a possibility that one may get rid of this property 
by imposing fine tuning relations
between the coefficients in the non-linear generalization of
the Fierz--Pauli 
mass term, order by order in non-linearity; however, these relations are not
protected by any obvious symmetry.
Thus, among others,
superluminality is a problematic
property to worry about when attempting to construct
infrared modifications of  General Relativity.

\end{abstract}

\section{ Introduction and summary}

Lorentz-invariant massive gravity in four dimensions --- the Fierz--Pauli theory ---
is of interest from the viewpoint of understanding the problems that may arise
when one attempts to modify General Relativity in the infrared domain.
Indeed, at the linearized level about Minkowski background, classical Fierz--Pauli
theory exhibits the van~Dam--Veltman--Zakharov discontinuity~\cite{vanDam:1970vg,Zakharov:1970cc}
and non-linearity at large distances from gravitating bodies~\cite{Vainshtein:1972sx},
whereas at the quantum level this theory becomes strongly coupled at unacceptably
low energies~\cite{Arkani-Hamed:2002sp}. In slightly curved backgrounds, the
Fierz--Pauli gravity and its non-linear generalizations necessarily contain an
extra, Boulware--Deser mode~\cite{Boulware:1973my}, over and beyond the five modes
of massive graviton; furhermore, one of the modes is always a ghost.
These properties of the Fierz--Pauli massive gravity are reviewed, e.g., in
Ref.~\cite{VRPT}.

In this note we point out that the Fierz--Pauli theory has yet another peculiar feature.
Namely, we find that, generically, 
there exist slighly curved backgrounds consistent with field equations,
in which some  of helicity-1 graviton modes are superluminal. We recall that these
modes are not pathological otherwise: they do not exhibit the vDVZ discontinuity,
Vainshtein non-linearity or particularly bad UV behavior at the 
quantum level\footnote{Helicity-0
(longitudinal) modes may be superluminal too, but they are  less interesting in our
context because of the other pathologies inherent in the longitudinal sector.}.
We check that the superluminal propagation occurs over large distances as compared
to the UV scale intrinsic in the Fierz--Pauli gravity, so this phenomenon shows up
within the domain of applicability of the effective low energy theory.

In this regard, the Fierz--Pauli massive gravity is analogous to the
Dvali--Gabadadze--Porrati model~\cite{Dvali:2000hr}, which also has a
superluminal mode in some legitimate backgrounds~\cite{Adams:2006sv}.
As discussed in Ref.~\cite{Adams:2006sv}, superluminal propagation 
signals that there is no UV completion of an effective low energy theory 
into quantum field theory
or perturbative string theory,
so the property we have observed is yet another feature to worry about.

Our analysis has a loophole, though. There remains a possibility that
the superluminal propagation of helicity-1 modes
may be avoided by imposing an infinite set of fine tuning relations on
the parameters in the non-linear generalization of the Fierz--Pauli mass
term, order by order in the degree of non-linearity. These relations,
however, are not protected by any obvious symmetry, so we do not find
this ``solution'' attractive.

It is worth stressing that we study here  massive gravities that have 
the Minkowski metric as a solution to the field equations. The situation 
may be less pathological in theories
whose ``natural'' background is different from Minkowski;
a well studied example is massive gravity about (anti)~de~Sitter 
space-time~\cite{Kogan:2000uy,Porrati:2000cp,Deser:2001pe,Deser:2001wx,Deser:2001xr,Porrati:2001db}.

Once Minkowski space-time is a solution to the field equations, the general mass term
is a polynomial in
$  ( g_{\mu \nu} - \eta_{\mu \nu})$.
The original Fierz--Pauli action is quadratic,
\begin{equation}
\label{FPaction}
S_{FP} = \frac{m_G^2}{64\pi G}\int d^4 x \left\{ -\eta^{\mu\lambda}\eta^{\nu\rho}
(g_{\mu\nu}-\eta_{\mu\nu})(g_{\lambda\rho}-\eta_{\lambda\rho}) + 
\left[\eta^{\mu\nu}(g_{\mu\nu}-\eta_{\mu\nu})   \right]^2  \right\} \; .
\end{equation}
Our purpose is to evaluate the speed of helicity-1 modes of metric perturbations
in nearly Minkowski backgrounds. So, we will make use of the perturbation theory
in $    (\overline{g}_{\mu \nu} - \eta_{\mu \nu}) $,
where $\overline{g}_{\mu \nu} $ is the background metric.
To this end, we will have to generalize the above expression by adding cubic, and
then higher-order terms. It is at this stage that the fine tuning ``solution'' 
of the superluminality problem shows up.

We begin with studying in section~\ref{sec:cosmo} 
the Fierz--Pauli theory and its generalizations
in cosmological backgrounds of small space-time curvature.
In these backgrounds, the metric perturbations decompose into the tensor,
vector and scalar modes with respect to spatial rotations.
To set the stage, we consider in section~\ref{sec:cosmo-FP} 
a particular form of the mass term 
(\ref{FPaction})
and 
find that the vector perturbations in this case are superluminal
in fairly general backgrounds. By itself, this is not a particularly strong
result, however.
The degree of superluminality 
$(c^2 -1)$, where $c$ is the speed of the propagation of the vector modes of metric
perturbations, is
proportional to the deviation of the background from Minkowski space-time,
 $(\overline{g}_{\mu \nu} - \eta_{\mu \nu})$.
This implies that the contributions to  $(c^2 -1)$ due to 
$\mathcal O\left[(g_{\mu \nu} - \eta_{\mu \nu})^3\right]$-terms in the action
are of the same order. We consider the cubic terms
in section~\ref{sec:cosmo-cubic-vector}  and find
that to the first order in  $(\overline{g}_{\mu \nu} - \eta_{\mu \nu})$,
the total $(c^2 -1)$ may be made equal to zero 
by fine tuning a parameter
in the cubic action. 
To see that similar fine tuning is required 
in higher  orders in
$(\overline{g}_{\mu \nu} - \eta_{\mu \nu})$, we have to study the action 
containing higher orders
in  $(g_{\mu \nu} - \eta_{\mu \nu})$. 

Proceeding with cosmological backgrounds is technically challenging at this stage.
Instead, we make use of the St\"uckelberg formalism. 
We consider backgrounds of a special form and gravitons traveling along a
particular direction. This enables us 
to separate helicity-1 modes from other metric perturbations
in a consistent way.
In section~\ref{sec:stuck-cub}
we cross check by rederiving, within this formalism,
the fine tuning relation ensuring that $c^2=1$ at the linear order in
$(\overline{g}_{\mu \nu} - \eta_{\mu \nu})$. In section~\ref{sec:stuck-quart} 
we show explicitly that for the backgrounds we consider,
the requirement that helicity-1 modes do not propagate in superluminal way
 gives rise, at the quadratic order in
$(\overline{g}_{\mu \nu} - \eta_{\mu \nu})$, to further fine tuning relations,
now involving the coefficients in the fourth-order action. 
It is then straightforward to convince oneself 
that fine tuning proliferates to higher orders. We have found no arguments 
suggesting that our fine tuning relations are sufficient to avoid 
superluminal propagation of helicity-1 modes in arbitrary backgrounds;
neither have we  found backgrounds that would rule out the fine-tuning 
``solution'' to the problem of superluminality in the helicity-1 sector.

\section{Cosmological backgrounds}
\label{sec:cosmo}

The theory we consider in this paper is 
the Fierz-Pauli model 
whose action is
\[
S = S_{GR} + S_m
\]
where $S_m$ is the mass term including the quadratic part (\ref{FPaction})
and possible higher order terms, while $S_{GR}$ is the action of General Relativity.
In the further analysis we set $16\pi G = 1$.
Our purpose in this section
is to study the propagation of gravitons in the spatially flat
cosmological backgrounds, in the regime when spatial momenta and frequencies
are much higher than the graviton mass.
In the coordinate frame where $\eta_{\mu\nu}$ in the mass term is 
$\diag[1,-1,-1,-1] $, the spatially flat FRW metric is
\be
ds^2 = n^2(t) a^2(t) dt^2 - a^2 (t) d{\bf x}^2
\label{c-bckgnd}
\ee
In what follows, it is convenient to use conformal time $\eta$
related to time
$t$ by
\[
  d\eta = n(t) dt
\]
In the coordinates $(\eta, {\bf x})$, the speed of light is equal to 1.
We will consider nearly flat metric, for which $\delta n = n-1$ and $\delta a=
a-1$ are small.

The field equations for the background, written
in terms of conformal time, read
\begin{equation}
\label{FPepsilons}
\left( \frac{a'}{a}\right) ^2 \equiv \mathcal H^2 = \epsilon_0,~~~
2\frac{a''}{a} - \mathcal H^2 = \epsilon_s
\end{equation}
where prime denotes $d/d\eta$,
the contributions $ \epsilon_0 \;, \; \epsilon_s$ 
are due to the mass term and are of order
\[
  \epsilon_0 \;, \; \epsilon_s = m_G^2 \left[ \mathcal O(a-1) + \mathcal O(n-1)\right]
\]
The lowest order terms come from the quadratic part (\ref{FPaction});
their explicit expressions are
\begin{equation}
\label{epsExplicit}
\epsilon_0 = -\frac{m_G^2 n}{2}(a^2 - 1),~~~
\epsilon_s = -\frac{m_G^2}{2n}(2(a^2 - 1) + (a^2 n^2 - 1))
\end{equation}
where the expansion up to the first order in $\delta a$ and $\delta n$ is understood.

Now, we can freely choose $\delta n$ and $\delta a$ at a given moment of time.
Then these variables will change in time at low rate determined by the value
of $m_G$, which we take small. We will consider the time scales
shorter than $m_G^{-1}$, so we treat $\delta n$ and $\delta a$
as constants.
It is worth noting that 
the consistency of the system
(\ref{FPepsilons}), i.e.,  covariant conservation of the effective
energy-momentum tensor coming from the mass terms, implies that $n(\eta)$
obeys the equation 
\be
\frac{\d \epsilon_0}{\d n} n^\prime = -  \frac{\d \epsilon_0}{\d a}  a^\prime
+ {\cal H} (\epsilon_s - \epsilon_0) \; .
\label{eq-n}
\ee
However, this equation determines the  evolution of $\delta n$ and
does not prohibit one to choose arbitrarily the value of $\delta n$ at a given
moment of time. So, for our purposes we can indeed treat $\delta n$ and $\delta a$ as
independent free parameters of the background.

Let us now include metric perturbations, and write the perturbed metric in
the following form,
\[
ds^2 = a^2(t)\left[ n^2(t)(1 + h_{00})dt^2 + 
2 h_{0i} n(t) dt dx^i +(-\delta_{i k} + h_{i k})dx^i dx^k   \right] \; .
\]
Due to the $O(3)$ symmetry of the FRW metric one can study 
different helicity sectors separately.
We will be interested in
vector (helicity-1) perturbations, for which, using the standard notation, we write
\bea
h_{i j} = \pd_i W_j + \pd_j W_i,~~~\pd_i W_i = 0,
\nonumber \\
~~~h_{00} = 0, ~~~h_{0i} = U_i,~~~ \pd_i U_i = 0
\nonumber
\eea
The question is whether or not the propagating vector modes are superluminal
for some choice of $\delta a$ and $\delta n$.

\subsection{Quadratic mass term only}
\label{sec:cosmo-FP} 

To warm up,
let us consider the particular form of the graviton mass term
given by (\ref{FPaction}), without cubic term. 
The complete quadratic action for vector
perturbations in the background obeying the equations of motion
(\ref{FPepsilons}) is
\bea
\left.S_{tot}^{(2)}\right|_{E.O.M.} = &&\frac{1}{2}\int a^2 d\eta d^3 x
 \left[\frac{}{}\right. 2U_i\Delta W_i' - U_i\Delta U_i + W_i''\Delta W_i + 
2\mathcal H W_i'\Delta W_i 
+ \left.6\epsilon_0 U_iU_i + 2\epsilon_s W_i\Delta W_i \frac{}{} \right] 
\nonumber \\
%
&&+ \int d^3 x dt 
\left( \frac{\delta m^2}{2}U_iU_i + \frac{\delta M^2}{2} W_i\Delta W_i \right)
\nonumber 
\eea
where $\Delta = \d_i \d_i$.
The last term in 
the above expression comes 
directly from the mass term in the action. In the case (\ref{FPaction})
the explicit expressions for $\delta m^2$, $\delta M^2$ to the first order
in $\delta n$, $\delta a$ are
\[
\delta m^2 = m_G^2 (1 + 4 \delta a + 2 \delta n),~~~ \delta M^2 = m_G^2
(1+ 4 \delta  a)
\]
It is convenient to define
\begin{equation}
M^2 \equiv 4\epsilon_s  + \frac{2}{a^2 n}\delta M^2,~~~m^2 \equiv 12\epsilon_0 + \frac{2}{a^2 n}\delta m^2
\nonumber
\end{equation}
Then the linear equations for metric perturbations are
\[
\left\{
\begin{aligned}
M^2\Delta W - 4\mathcal H\Delta U - 2\Delta U' + 2\Delta W'' + 4\mathcal H\Delta W' = 0 \\
m^2 U + 2\Delta W' - 2\Delta U = 0 \\
\end{aligned}
\right.
\]
We now concentrate on frequency and momenta exceeding $H_0$ and $m_G$,
and make the Fourier transformation 
$\frac{\d}{\d \eta} \to i\omega,~\Delta \to -p^2$.
In this limit we obtain the following dispersion relation for the vector perturbations,
\begin{equation}
\label{FPgenResult}
c^2 \equiv \frac{\omega^2}{p^2} = \frac{M^2}{m^2}
\end{equation}
To the first order in $\delta n$, $\delta a$ we find
\begin{equation}
\label{FPresult}
c^2 - 1 = -4\delta n
\end{equation}
This implies that for any background with $\delta n < 0 $, 
the speed of vector perturbations exceeds the speed of light.

Importantly, $c^2 -1$ as given by (\ref{FPresult}) is of the zeroth order
in $m_G^2$. On the other hand, $\delta n^\prime$ is of the first
order in this parameter. So, at small  $m_G^2$ there is
enough time for the vector perturbations to propagate, in superluminal way,
before the space-time metric changes. In detail, 
the metric does not change during the time interval 
$\tau \sim m_G^{-1} \cdot |\delta a|^{1/2}$,
where the factor $|\delta a|^{1/2}$ is 
obtained by inspecting eqs.~(\ref{epsExplicit})
and (\ref{eq-n}). 
The distance the vector
perturbations advance light in this time interval is
$L = \delta c \cdot \tau \sim m_G^{-1} \cdot |\delta a|^{1/2} |\delta n|$, 
which for not very small
$\delta a$ and $\delta n$ is much greater than
the strong coupling scale~\cite{Arkani-Hamed:2002sp}
inherent in the Fierz--Pauli theory,
$\Lambda_{UV}^{-1} = (m_G^4 M_{Pl})^{-1/5}$. Hence,
the superluminal propagation is not hidden below the UV distance
scale.
This reasoning applies word by word
to massive gravities
with more general 
actions, to be discussed below. The effect of 
the superluminal propagation we analyze  in this paper
does occur within the domain of applicability of the
effective field theory.

\subsection{The cubic action}
\label{sec:cosmo-cubic-vector}

To the linear order in $\delta n$, $\delta a$, the speed of vector
perturbations may receive contribution from cubic in $(g_{\mu\nu} -\eta_{\mu \nu})$ 
part of the mass term.
One may thus wonder whether 
superluminality may be eliminated by the 
proper choice of the parameters in this part of the action. 
To examine this issue,
 let us calculate the linear in 
$\delta a,~\delta n$ contribution to  $c^2$ for general cubic term. 
As before, the mass term in the action
is assumed to be Lorentz-invariant, 
contain no derivatives and  
have  flat metric as a solution to the massive gravity equations. 
Its quadratic  part is of the Fierz--Pauli structure.
The general expression for the cubic part is
\begin{equation}
\label{cubicLagrangian}
S_m^{(3)} = \int~d^4x~\left(
\mathcal C\cdot A_1^3 + \mathcal E\cdot A_2 A_1 + \mathcal F\cdot A_3 \right) \; ,
\end{equation}
where
\bea
A_1 &=& \eta^{\mu\nu}(g_{\mu\nu}-\eta_{\mu\nu}),~~~
A_2 = \eta^{\mu\lambda}\eta^{\nu\rho}(g_{\mu\nu}-\eta_{\mu\nu})(g_{\lambda\rho}-\eta_{\lambda\rho}) \; ,
\nonumber\\
& & ~~~~A_3 = \eta^{\mu\lambda}\eta^{\rho\sigma}\eta^{\nu\tau}(g_{\mu\nu}-\eta_{\mu\nu})
(g_{\lambda\rho}-\eta_{\lambda\rho})(g_{\sigma\tau}-\eta_{\sigma\tau})
\label{defAs}
\eea
with $\mathcal C,~\mathcal E,~\mathcal F$ being 
arbitrary coefficients. The total mass part of the action is
the Fierz-Pauli term (\ref{FPaction}) plus $S_m^{(3)}$.

The expressions (\ref{FPepsilons}) -- (\ref{FPgenResult}) 
are still valid, but $\delta m$ and $\delta M$
are to be recalculated. The linearized expressions are
\bea
\delta m^2 &=& m_G^2(1 + 4\delta a + 2\delta n)
- 8\mathcal E (4\delta a + \delta n) - 12\mathcal F(2\delta a + \delta n)
\nonumber \\
\delta M^2 &=& m_G^2(1 + 4\delta a) 
- 8\mathcal E (4\delta a + \delta n) - 24\mathcal F\delta a 
\nonumber
\eea
The previous result (\ref{FPresult})
then modifies to 
\begin{equation}
\label{vectorResult}
c^2 - 1 = \frac {3\mathcal F-m_G^2}{m_G^2}\cdot 4\delta n
\end{equation}
Thus,  to avoid superluminality at the first order in $\delta a$,
$\delta n$, one has to impose the fine tuning
relation
\be
3\mathcal F = m_G^2
\label{tuning}
\ee
We will reproduce this fine tuning relation in section~\ref{sec:stuck-cub} in the
St\"uckelberg approach, and in section~\ref{sec:stuck-quart} we 
proceed to higher order terms.

\section{The St\"uckelberg approach }
\subsection{Cubic order fine tuning  revisited}
\label{sec:stuck-cub}

To continue the analysis, it is convenient to use the St\"uckelberg formalism, wich will also
allow us to study backgrounds other than cosmological. 
Let us write the metric 
as follows,
\be
g_{\mu\nu} = \eta_{\mu\nu} + \overline h_{\mu\nu} + h_{\mu\nu} \; ,
\label{genmetric}
\ee
 where
the first term is the Minkowski metric, the second term
corresponds to non-trivial background and is assumed to
be small, and the third one describes perturbations about this background.

To perform the St\"uckelberg
analysis, one enlarges the set of fields in the theory by
introducing new fields $\xi^\mu$ and $\tilde{g}_{\mu \nu}$
in the way dictated by
the gauge symmetry of General 
Relativity\cite{Arkani-Hamed:2002sp,Dubovsky:2005dw,Creminelli:2005qk,Deffayet:2005ys},
see also Ref.~\cite{VRPT} for detailed discussion,
\begin{equation}
  g_{\mu \nu}(x) =
\tilde{g}_{\mu \nu}(x+\xi) + \d_\mu \xi^\lambda \; \tilde{g}_{\nu \lambda}(x+\xi)
 + \d_\nu \xi^\lambda \; \tilde{g}_{\mu \lambda}(x+\xi) +
   \d_\mu \xi^\lambda \; \d_\nu \xi^\rho \; \tilde{g}_{\lambda \rho}(x+\xi) \; .
\nonumber
\end{equation}
We are going to consider high graviton momenta and slowly varying backgrounds,
so we neglect the derivatives of $\overline{h}_{\mu \nu}$. 
Once the field  $\tilde{g}_{\mu \nu} (x)$ is gauge fixed, mixing 
between its fluctuations and the field $\xi^\mu$ becomes
irrelevant at high enough momenta, and we are left with the theory of  one vector 
field $\xi^\mu$ determining the interesting part of metric perturbations via
\begin{equation}
\label{stuckFields}
h_{\mu\nu} = 
\pd_\mu \xi_\nu +\pd_\nu\xi_\mu +\pd_\mu \xi^{\lambda}\overline{h}_{\lambda\nu}
 +\pd_\nu \xi^{\lambda}\overline{h}_{\lambda\mu}
 + \pd _\mu\xi^\lambda \pd_\nu \xi_\lambda
+ \overline{h}_{\lambda\rho}\pd_\mu\xi^{\lambda}\pd_\nu\xi^{\rho} \; .
\end{equation}
 Throughout this paper indices are raised and lowered with the Minkowski metric.
Note 
that, nevertheless,
 this expression is exact for slowly varying backgrounds, i.e., no expansion
in $\overline{h}_{\mu \nu}$ has been made yet.

The field $\xi^\mu$ does not enter the Einstein--Hilbert part of the action,
and obtains kinetic term from the mass term in the original action. To evaluate its speed, 
we study the action to the quadratic order in $\xi^\mu$.
We begin with the linear order in $\overline{h}_{\mu \nu}$; in this way we are going to
reproduce the fine tuning relation (\ref{tuning}) in the St\"uckelberg formalism.
Hence, we need the mass term to the cubic order in $(g_{\mu \nu} - \eta_{\mu \nu})$,
\[
S_m^{(2+3)} 
= S_{FP} + S_m^{(3)} = \int~d^4x \left[\mathcal A\cdot (-A_1^2 + A_2) + \mathcal C\cdot A_1^3
 + \mathcal E\cdot A_2 A_1 + \mathcal F\cdot A_3 \right] \; ,
\]
where 
\be
\mathcal A \equiv -m_G^2/4 \; ,
\label{A-definition}
\ee
and the combinations $A_1$, $A_2$, $A_3$ are given by (\ref{defAs}). According to the 
above discussion,
we will need the terms in this action 
that are quadratic in the St\"uckelberg field $\xi^{\mu}$ and zero and first order
 in $\overline{h}_{\mu\nu}$. Note that these terms 
appear both due to the quadratic  and cubic terms in the action 
and due to the non-linearity of the gauge transformation (\ref{stuckFields}).

Making use of (\ref{genmetric}) and (\ref{stuckFields}) one obtains
$S_m^{(2+3)} = \int~d^4x~ \mathcal{L}_m^{(2+3)} $, with
\bea
\mathcal{L}_m^{(2+3)} (\overline h, \xi) = && \left\{ 
- 2\mathcal A \, [ (\pd_{\mu}\xi^{\mu})^2 + \xi_{\mu}\Box\xi^{\mu}] - 
(2\mathcal A+3\mathcal F)\, 
\xi_{\lambda}\overline h^{\mu\nu}\pd_\mu\pd_\nu \xi^{\lambda}
\right.
\nonumber \\
&& \left. 
+ 2(4\mathcal E+3\mathcal F - 2\mathcal A)\, \pd_{\lambda}\xi^{\lambda} \cdot
\overline h^{\mu\nu}\pd_{\mu}\xi_{\nu} \right.
\nonumber \\
&& \left.
+ 2(\mathcal A-\mathcal E) \, \overline h \xi_{\mu}\Box\xi^{\mu}
+ 2(\mathcal E+6\mathcal C)\, \overline h (\pd_{\mu}\xi^{\mu})^2 
- (3\mathcal F + 4\mathcal A)\, \overline h_{\mu\nu}\xi^\mu\Box \xi^\nu \right\}\; ,
\label{Stuckac-3}
\eea
where 
\[
\overline h \equiv \overline h_{\mu\nu}\eta^{\mu\nu} \; .
\]
Let us consider the frame
in which the Minkowski metric entering the mass terms has the standard form
$\eta_{\mu \nu} = \mbox{diag} [1,-1,-1,-1]$ and study the backgrounds whose
space-time metric is 
diagonal in this frame,
$\overline h_{\mu\nu} = \diag[\overline h_{00},
\overline h_{11},\overline h_{22},\overline h_{33}]$,
so that
\be
ds^2 = (1+\overline h_{00})dt^2 ~+~ \sum\limits_i (-1+\overline h_{ii})(dx^i)^2 \; .
\label{diag-metric}
\ee
For time-independent $\overline{h}_{\mu \nu}$ this metric is a solution to the Einstein equations,
so in massive theory the derivatives of $\overline{h}_{\mu \nu}$ are proportional to $m_G^2$.
Hence, it is legitimate to consider this metric as slowly varying. This justifies
the use of the expression (\ref{stuckFields}), and also enables us to neglect the derivatives
of $\overline{h}_{\mu \nu}$ when solving the equations for $\xi^\mu$.

Let us further specify to
 gravitons propagating along the 1-st axis, whose momenta are
\[
P_0 = \omega,~~~ P_i = p\delta_{\, i\, 1}
\]
In this case the action (\ref{Stuckac-3}) is a sum of  three independent
parts: the action that involves the field
$\xi^2$ only, the action for $\xi^3$ and the action for the pair $(\xi^0, \xi^1)$.
An obvious reason for this decoupling
is that the matrix $\overline{h}_{\mu \nu}$ does not mix, say, $\xi^2$ with
other components of $\xi^\mu$, while $\d_\mu \xi^\mu$ involves $\xi^0$ and $\xi^1$ only.
Hence, the transverse polarizations $\xi^2$ and $\xi^3$ decouple from each other and from
longitudinal polarizations. These are precisely the (linear combinations of)
helicity~$\pm 1$ modes we are interested in. Their
dispersion relations are readily calculated.
One finds that the dispersion relation for both $\xi^2$ and $\xi^3$  to the linear order 
in $\overline{h}_{\mu \nu}$ is
\[
   \omega^2 = \left[ 1 - \frac{2 \mathcal{A} + 3 \mathcal{F}}{2 \mathcal{A}} (\overline{h}_{00}
+ \overline{h}_{11})\right] \cdot  p^2 \; .
\]
The physical speed of these modes 
 (recall that they propagate along the 1-st axis) is
\be
c^2 = \frac{1}{n^2} \frac{\omega^2}{p^2} \; ,
\label{physspeed}
\ee
where
\be
 n^2 
\equiv \frac{g_{00}}{|g_{11}|} = \frac{1+\overline h_{00}}{1-\overline h_{11}} 
\label{def-n2}
\ee
Hence to the 
linear order in $\overline{h}_{\mu \nu}$, the physical speed of helicity-1 gravitons is given by
\be
c^2 = 1 -\frac { (4\mathcal A+3\mathcal F)}{2 \mathcal A}  
(\overline{h}_{00} + \overline{h}_{11})
\label{3stuck-res}
\ee
where  $(\overline{h}_{00} + \overline{h}_{11})$
may have either sign. Recalling (\ref{A-definition}) and
having in mind the two forms of background metric
(\ref{c-bckgnd}) and (\ref{diag-metric}),
we see that the expression
(\ref{3stuck-res}) coincides with (\ref{vectorResult}).
In this way we recover the result of section \ref{sec:cosmo-cubic-vector}:
the condition for the absence of superluminal propagation of helicity-1
modes at the linear order in  $\overline{h}_{\mu \nu}$ is
$ 3\mathcal{F} = - 4 \mathcal{A} \equiv m_G^2$.

\subsection{Fine tuning at fourth order and beyond}
\label{sec:stuck-quart}

Let us now study the theory tuned according to (\ref{tuning}). At the first order in
$\overline{h}_{\mu \nu}$, helicity-1 gravitons in this theory propagate at high enough
momenta precisely with the speed of light. We will now see that at the second order,
superluminlity of helicity-1 gravitons generically
reappears, unless one imposes additional fine tuning relations.

Let us introduce the general 
fourth-order mass terms and  expand the action up 
to the order of $\mathcal O(\overline{h}^2\times h^2)$. 
The fourth order terms are
\[
S_m^{(4)} = \int~d^4x~\left[
\mathcal I\cdot (A_4) + \mathcal J\cdot (A_1)^4 + 
\mathcal K\cdot (A_1 A_3) + \mathcal L\cdot (A_1^2 A_2) + \mathcal M(A_2)^2 \right] \; .
\] 
Here, in analogy to  $A_{2,3}$ in (\ref{cubicLagrangian}), the irreducible term is
\[
A_4 = \eta^{\mu_2 \nu_1} \dots \eta^{\mu_1 \nu_4} \cdot
(g_{\mu_1 \nu_1} - \eta_{\mu_1 \nu_1}) \dots (g_{\mu_4 \nu_4} - \eta_{\mu_4 \nu_4})
\]
and $\mathcal I, \dots \mathcal M$ are arbitrary constants.
Plugging the St\"uckelberg decomposition (\ref{stuckFields})
into this action, and keeping the terms  of order $\mathcal O(\overline{h}^2\times h^2)$
in the complete action, we obtain the contribution to the quadratic
action of
the St\"uckelberg fields in the form
$
S_m^{(4)} = \int~d^4x~\mathcal{L}^{(4)}
$,
where
\be
\begin{aligned}
\mathcal{L}_m^{(4)}(\overline h,\xi) = 
&-(3\mathcal C + 2\mathcal L) \, (\overline h^2 \xi^\mu \Box \xi_\mu) - 
(\mathcal E + 4\mathcal  M)\, (\overline h_{\nu\rho} \overline{h}^{\nu \rho} \cdot
\xi^\mu \Box \xi_\mu)
- (2\mathcal E + 3\mathcal  K) \, (\overline h \xi^\lambda \overline h^{\mu\nu}
\pd_\mu\pd_\nu \xi_\lambda)\\
&- (3\mathcal F + 4\mathcal I)\, 
(\xi^\lambda \overline h^{\mu\rho}\overline h_{\rho}^\nu \pd_\mu\pd_\nu \xi_\lambda)
- (2\mathcal A + 6\mathcal F + 4\mathcal I)\, (\overline h_{\mu\rho}\overline h^{\rho\nu}\xi^\mu \Box \xi_\nu)\\
&+ 2(3\mathcal F + 4\mathcal E + 4\mathcal I + 6\mathcal K) \,
(\pd_\lambda\xi^\lambda \cdot \overline h^{\mu}_\nu \overline h^{\nu\rho}\pd_\mu\xi_\rho)
+ 2(12\mathcal J + \mathcal L)\, \overline h^2(\pd_\mu\xi^\mu)^2\\
&- (3\mathcal  K + 4\mathcal  E - 2\mathcal A)\, (\overline h \overline h_{\mu\nu}\xi^\mu \Box \xi^\nu)
+ (6\mathcal  K + 16\mathcal L + 24\mathcal C + 4\mathcal E)\, (\overline h 
\pd_\lambda\xi^\lambda \cdot \overline h^{\mu\nu}\pd_\mu\xi_\nu)\\
&+ 2(-\mathcal A + 3\mathcal F + 4\mathcal E + 8\mathcal M + 2\mathcal I)\,
(\overline h^{\mu\nu}\pd_\mu\xi_\nu)^2 
- (2\mathcal A + 6\mathcal F + 4\mathcal I) \, (\xi^\rho \overline h^{\mu\nu} \overline h_{\rho\lambda} \pd_\mu\pd_\nu 
\xi^\lambda )
\end{aligned}
\nonumber
\ee
The total Lagrangian is the sum of this term and (\ref{Stuckac-3}).
We continue to consider the diagonal background metric (\ref{diag-metric}) and gravitons 
propagating along the 1-st axis. The 
fields $\xi^2$ and $\xi^3$ again decouple, for the same reason
as in section \ref{sec:stuck-cub}. Making use of  (\ref{tuning}), we obtain the following
dispersion relation for the coordinate frequency  of
the mode $\xi^2$ to the second order in $\overline{h}_{\mu \nu}$
\bea
\omega^2 = && \left\{ 1 + (\overline{h}_{00} + \overline{h}_{11}) 
\phantom{\frac{}{}}
\right.
\nonumber\\
 && \left. + \left[ \overline{h}_{00} + \frac{3 \mathcal{A} - 2 \mathcal{I}}{\mathcal A} \overline{h}_{22} 
+ \frac{2 \mathcal{A} - 4 \mathcal{E}
- 3 \mathcal{K}}{2\mathcal A} \overline{h} + 
2\frac{\mathcal{A} - \mathcal{I}}{\mathcal A} (\overline{h}_{00} - 
\overline{h}_{11}) \right]  (\overline{h}_{00}
+ \overline{h}_{11}) \right\} \cdot p^2 \; .
\nonumber
\eea
Finally, expanding $n^{-2}$, defined according to (\ref{def-n2}), to the
second order in  $\overline{h}_{\mu \nu}$ we find that 
the physical speed (\ref{physspeed}) of the helicity-1 graviton with the
polarization
$\xi^2$  is given by
\be
c^2 
= 1 + \left[  \frac{3 \mathcal{A} - 2 \mathcal{I}}{\mathcal A} \overline{h}_{22} + 
\frac{2 \mathcal{A} - 4 \mathcal{E}
- 3 \mathcal{K}}{2\mathcal A} \overline{h} + 
\frac{3 \mathcal{A} -2\mathcal{I}}{\mathcal A} 
(\overline{h}_{00} - \overline{h}_{11}) \right] (\overline{h}_{00}
+ \overline{h}_{11})
\nonumber
\ee
Since the quantities $(\overline{h}_{00}+ \overline{h}_{11})$, 
$(\overline{h}_{00} - \overline{h}_{11})$, $\overline{h}_{22}$ 
and $\overline{h}$
are independent of each other, we see that 
to avoid the superluminal propagation,
one has to impose additional fine tuning 
relations 
\be
3 \mathcal{A} -2\mathcal{I} = 0 \; , \; \; \; \; \; 2 \mathcal{A} - 4 \mathcal{E}
- 3 \mathcal{K} = 0 \; .
\label{may7-1}
\ee
The same relations ensure that the mode $\xi^3$ is not superluminal in the
background we consider.

The analysis in this section reveals the following property:
the dispersion relation for helicity-1 gravitons involves the
coefficients of the irreducible higher order terms, which are
multiplied by sign-indefinite combinations of the background metric.
At the third order, the relevant coefficient is $\mathcal{F}$, the
irreducible term is $A_3$, and the combination of metric is
$(\overline{h}_{00}
+ \overline{h}_{11})$, while
at the fourth order, these are $\mathcal{I}$,  $A_4$, and 
$\overline{h}_{22} (\overline{h}_{00}
+ \overline{h}_{11})$, respectively. Unless these coefficients are fine tuned,
helicity-1 gravitons propagate in superluminal way in 
backgrounds with appropriate signs of the background metric coefficients
$\overline{h}_{\mu \nu}$. Indeed,  the relation (\ref{tuning}) contains  $\mathcal{F}$,
and
the first of the relations (\ref{may7-1})
contains  $\mathcal{I}$. 
One can check that this property holds at higher orders. Hence, the superluminal 
propagation reappears at higher orders unless at least one fine tuning relation
is imposed at each order in $(g_{\mu \nu} - \eta_{\mu \nu})$ (in fact,  our 
fourth-order analysis
shows that the number of fine tuning relations is larger than one).

There is no obvious symmetry behind the relations like (\ref{tuning}) and  (\ref{may7-1}).
Thus, helicity-1 gravitons propagate superluminally in the Fierz--Pauli massive gravity
unless this theory is heavily fine tuned. 

\vspace{0.5cm}

 This work
has been supported in part by Russian Foundation for Basic
Research, grant 08-02-00473.


\begin{thebibliography}{10}

\bibitem{vanDam:1970vg}
H.~van Dam and M.~J.~G. Veltman,
\newblock Nucl. Phys. {\bf B22}, 397 (1970).

\bibitem{Zakharov:1970cc}
V.~I. Zakharov,
\newblock JETP Lett. {\bf 12}, 312 (1970).

\bibitem{Vainshtein:1972sx}
A.~I. Vainshtein,
\newblock Phys. Lett. {\bf B39}, 393 (1972).

\bibitem{Arkani-Hamed:2002sp}
N.~Arkani-Hamed, H.~Georgi and M.~D. Schwartz,
\newblock Ann. Phys. {\bf 305}, 96 (2003), [hep-th/0210184].

\bibitem{Boulware:1973my}
D.~G. Boulware and S.~Deser,
\newblock Phys. Rev. {\bf D6}, 3368 (1972).

\bibitem{VRPT}
V.~A. Rubakov and P.~G. Tinyakov,
\newblock arXiv:0802.4379 [hep-th].

\bibitem{Dvali:2000hr}
G.~R. Dvali, G.~Gabadadze and M.~Porrati,
\newblock Phys. Lett. {\bf B485}, 208 (2000), [hep-th/0005016].

\bibitem{Adams:2006sv}
A.~Adams, N.~Arkani-Hamed, S.~Dubovsky, A.~Nicolis and R.~Rattazzi,
\newblock JHEP {\bf 10}, 014 (2006), [hep-th/0602178].

\bibitem{Kogan:2000uy}
I.~I. Kogan, S.~Mouslopoulos and A.~Papazoglou,
\newblock Phys. Lett. {\bf B503}, 173 (2001), [hep-th/0011138].

\bibitem{Porrati:2000cp}
M.~Porrati,
\newblock Phys. Lett. {\bf B498}, 92 (2001), [hep-th/0011152].

\bibitem{Deser:2001pe}
S.~Deser and A.~Waldron,
\newblock Phys. Rev. Lett. {\bf 87}, 031601 (2001), [hep-th/0102166].

\bibitem{Deser:2001wx}
S.~Deser and A.~Waldron,
\newblock Phys. Lett. {\bf B508}, 347 (2001), [hep-th/0103255].

\bibitem{Deser:2001xr}
S.~Deser and A.~Waldron,
\newblock Phys. Lett. {\bf B513}, 137 (2001), [hep-th/0105181].

\bibitem{Porrati:2001db}
M.~Porrati,
\newblock JHEP {\bf 04}, 058 (2002), [hep-th/0112166].

\bibitem{Dubovsky:2005dw}
S.~L. Dubovsky, P.~G. Tinyakov and I.~I. Tkachev,
\newblock Phys. Rev. {\bf D72}, 084011 (2005), [hep-th/0504067].

\bibitem{Creminelli:2005qk}
P.~Creminelli, A.~Nicolis, M.~Papucci and E.~Trincherini,
\newblock JHEP {\bf 09}, 003 (2005), [hep-th/0505147].

\bibitem{Deffayet:2005ys}
C.~Deffayet and J.-W. Rombouts,
\newblock Phys. Rev. {\bf D72}, 044003 (2005), [gr-qc/0505134].

\end{thebibliography}

\end{document}